# Color Intensity Projections
## A simple way to display changes in astronomical images

Keith S Cover[1], Frank J. Lagerwaard[1], Suresh Senan[1]
VU University Medical Center, Amsterdam[1]

Corresponding author:
Keith S Cover, PhD
VU University Medical Center
Post Box 7057
1007 MB Amsterdam
The Netherlands
Email: Keith@kscover.ca

## Abstract

To detect changes in repeated astronomical images of the same field of view (FOV), a common practice is to stroboscopically switch between the images. Using this method, objects that are changing in location or intensity between images are easier to see because they are constantly changing. A novel display method, called arrival time color intensity projections (CIPs), is presented that combines any number of grayscale images into a single color image on a pixel by pixel basis. Any values that are unchanged over the grayscale images look the same in the color image. However, pixels that change over the grayscale image have a color saturation that increases with the amount of change and a hue that corresponds to the timing of the changes. Thus objects moving in the grayscale images change from red to green to blue as they move across the color image. Consequently, moving objects are easier to detect and assess on the color image than on the grayscale images. A sequence of images of a comet plunging into the sun taken by the SOHO satellite (NASA/ESA) and Hubble Space Telescope images of a trans-Neptunian object (TNO) are used to demonstrate the method.

**Subject headings**: methods: data analysis, techniques: image processing, comets: general

# 1 INTRODUCTION

Detection of changes in a sequence of images of the same region of the sky is widely used in astronomy to detect comets, asteroids, supernova, variable stars and other objects. While there are several fully automated methods to detect these changes (Jewitt et al. 1998, Gladman 1998, Allen et al. 2001, Petit et al 2004), in many cases visual inspection for these changes is still used both for detection and for reviewing the results of changes detected automatically.

A commonly used method of visual inspection is to stroboscopically switch between images. Alternately, if the comparison of only 2 or 3 images is required, the images are sometimes displayed in the red, green and blue image planes (Buie et al. 2003). In the resulting color image, constant objects appear grayscale while changing objects appear in color. Buie et al. report the detection of changes by visual inspection to be much faster for the three color method than the stroboscopic method. They also report the three color method also allows the detection of some objects that have gone undetected by fully automated methods. However, a major limitation of the three color method is that it is limited to only 3 images because there are only 3 primary colors. Also, a major limitation of any method that depends on visual inspection is that it requires human intervention with all it complications including human error.

The method of color intensity projections (CIPs) was originally developed by the authors for applications in medical imaging but it was apparent from the start that the method may find useful applications in other fields including astronomy. The first publication of CIPs demonstrated its use in presenting the motion of lung tumors and normal organs during respiration (Cover et al. 2006). In the first version of CIPs, called percentage time CIPs, the hue of the color is used to encode the percentage of time there is tissue in a pixel. The second publication used the arrival time version of CIPs to visualize the flow of blood into and out of the brain using an X-ray contrast agent (Cover et al. 2007). The arrival time version of CIPs is calculated by adding a preprocessing step to the percentage time method. All the details of the arrival time CIPs method are described in the references but are also included in this paper for convenience.

While the primary mission of the SOHO satellite is to constantly monitor the sun, it has a long history of observing comets both grazing the sun and plunging into it (Sekanina 2002; Biesecker 2002). In particular, the LASCO C3 coronagraph of SOHO (Morrill et al. 2006) has observed many of these comets. With its field of view extending from 3.8 to 30 solar radii and images acquired dozens of times a day, the C3 coronagraph provides ideal data for applying CIPs.

The detection of trans-Neptunian objects (TNO) rely on the changes from one astronomical image to the next. Hubble Space Telescope (HST) (NASA/ESA) images of the binary TNO 1998 WW31 are also used to demonstrated the CIPs display method (Veillet et al. 2002)

## 2 METHOD

A total of 18 grayscale images of a comet plunging into the sun that were acquired by the SOHO LASCO C3 coronagraph during November 2nd and 3rd, 2006 are used to demonstrate the arrival time CIPs display technique. The images were cropped to 1014 by 690 pixels for processing and display purposes. In addition, six grayscale images of the binary TNO 1998 WW31 acquired by the Hubble Space Telescope between July 2001 and February 2002 where also used for demonstration. The HST images were cropped to 50 by 50 pixels for processing.

Grayscale images are characterized by a single value per pixel. While they are often displayed in grayscale they can also be presented in false color. Color images are defined by 3 values per pixel. The values may correspond to the intensity of the red, green and blue components or some other numerical representation of color.

The first step in calculating an arrival time CIP image is to calculate a new image sequence from the original time ordered sequence. The new sequence, which will be referred to as the cumulative maximum intensity projection (cumulative MaxIP), is simple to calculate. The first image of the cumulative MaxIP sequence is set equal to the first image of the original image sequence. The second image of the cumulative MaxIP sequence is set equal to the pixel by pixel maximum of the first two images of the original image sequence. The third image of the cumulative MaxIP sequence is set equal to the pixel by pixel maximum of the first three images of the original image sequence. Following the same pattern, all the images of the cumulative MaxIP sequence can be calculated.

The arrival time CIPs is then calculated by applying the percentage time CIPs method to the cumulative MaxIP sequence. The first step is to scale the pixel values of the cumulative MaxIP sequence to values between 0 and 1. This new image sequence, the normalized cumulative MaxIP sequence, is calculated by dividing each pixel in each cumulative MaxIP image by the maximum value of all pixels in all images. The second step is to calculate, on a pixel-by-pixel basis, the maximum, mean and minimum of the normalized cumulative MaxIP sequence. The maximum ($P_{MAX}$), mean ($P_{MEAN}$) and minimum images ($P_{MIN}$) are then combined, on a pixel by pixel basis, using the following equations to get the hue, saturation and brightness (HSB) of the composite color image:

$$\text{Brightness} = P_{MAX}$$
$$\text{Saturation} = (P_{MAX} - P_{MIN})/P_{MAX} \tag{1}$$
$$\text{Hue} = 2/3 * (P_{MEAN} - P_{MIN})/(P_{MAX} - P_{MIN})$$

The HSB for each pixel is then converted to the standard red, green and blue (RGB) presentation of color, using standard subroutines (Arnold et al. 2005). The hue is multiplied by 2/3 so there is a clear difference between the hue corresponding to the earliest and latest arrival times.

Calculating a CIP for the 18 SOHO images with 1014 by 690 pixels took less than a second to calculate on an AMD Duron running at 1GHz. If additional processing speed is needed for a very large data set, the algorithm can be easily parallelized because each pixel is calculated independently.

## 3 RESULTS AND DISCUSSION

The upper 4 panels of Fig. 1 (a-d) show 4 of 18 images of a comet plunging into the sun taken by the C3 coronagraph. Structures common to all the images are 1) the occulting disk in the center of each image that blocks out the sun and 2) jets of gas emerging radially out from the sun in several directions.

The comet is clearly visible on panels (b) and (c) to the lower right of the sun. Extrapolating the motion of the comet makes the comet easier to find on panels (a) and (d). Closer examination of the images also shows the apparent motion of the stars from the left to the right on the images. The apparent motion of the stars is actually due to the motion of the SOHO satellite, along with the earth, in orbit around the sun.

As CIPs images present pixels that are unchanged in all the grayscale images as a shade of gray in the color image, it is easy to pick out in the CIP image, panel (e), the structures that are unchanged over the grayscale images. These are the occulting disk and the jets of gas.

The moving objects in the CIP image, such as stars, planets and the comet, change from red to green to blue as they move across the FOV, as indicated by the color scale in the upper left corner of the CIP. Careful examination of the CIP will show the very first occurrence of an object is actually a shade of gray. This is another characteristic of arrival time CIPs.

The comet plunging into the sun can be clearly seen to the lower left of the sun in the CIP image. The uneven spacing between the various hues of the comet are due to the uneven timing of the C3 images. The movement of the background stars from the left to the right of the CIP standout because their tracks are all the same length.

Two planets can also be seen in the CIPs image. An object brighter than most of the stars and with a track less than half the length of the stars can be seen to the right of the sun. It is the planet Mars. The very bright object just to the upper left of the sun is also a planet. The planet, which is Venus, can be easily seen to be moving in the opposite direction to the stars.

Some objects only appear in a single hue, such as the yellow object near the base of the rod supporting the occulting disk. The single hue indicates a one time event that only appears on a single image, and the hue itself is indicative for the timing of this one time event. Artifacts such as cosmic rays can be discerned by this characteristic.

Through the example of SOHO images of a comet plunging into the sun, CIPs has been demonstrated as a simple and fast method for the detection of changes in astronomical images. While CIPs has been demonstrated for medical imaging and astronomical applications, there are likely a wide range of other applications where it may be applied.

The relative motion of the binary TNO in Fig. 2 is clearly shown by the CIP. One of the binary objects appears stationary as the grayscale images were shifted to align it on all 6 grayscale images. The orbit and direction of motion of the second object relative to the first is clearly apparent from the CIP. As the grayscale images were not evenly spaced in time, the hue gives the order rather than the precise timing of the object in the grayscale images.

One of the strengths of CIPs is that many different types of information can be encoded in the hue. As mentioned in the Methods section, the hue can be used to represent the percentage time an object spends at a particular location rather than the time it arrives at the location. Another possible modification is to make the first occurrence of an object have a red hue rather than the grayscale of arrival time CIPs. This can be accomplished by modifying the percentage time CIPs so that the hue is determined by the time of the peak signal in each pixel.

Another variation of CIPs that may be useful in the detection and visualization of variable stars is introducing artificial motion to the images. This can be accomplished by shifting each grayscale image by a pixel or two relative to the previous image before calculating the CIPs. Thus, fixed stars will appear to move in a manner similar to SOHO images as shown in Fig. 1. The artificial motion will make the variation of the intensity of a star over time more apparent. If a star field is crowded, circular or spiral patterns of artificial motion might be worth considering to reduce overlaps.

Whichever version of CIPs are being considered for an application, it would be best to apply the various versions of CIPs to a variety of data sets to find which variation performs the best for each application.

## ACKNOWLEDGEMENTS

This work was funded by the VU University Medical Center, Amsterdam. The VU University Medical Center is pursuing a patent on color intensity projections on behalf of the authors.

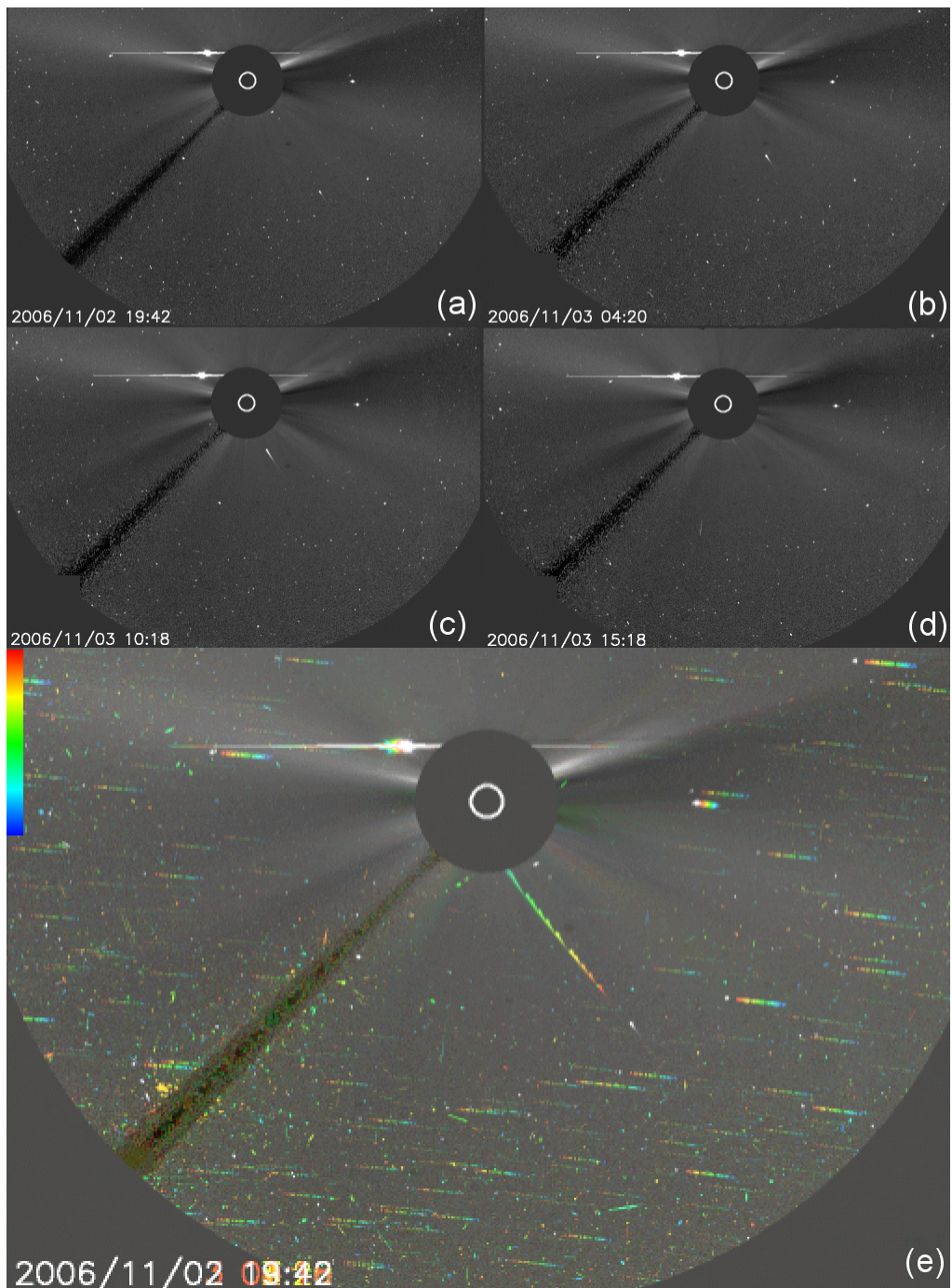

Figure 1. Images from the SOHO LASCO C3 coronagraph showing a comet plunging into the sun. Panels (a) - (d) show the standard presentation for 4 of the 18 grayscale images acquired. Panel (e) shows the arrival time CIP of the 18 component images. The color bar in the upper right corner of the CIP gives the timing of the different hues.

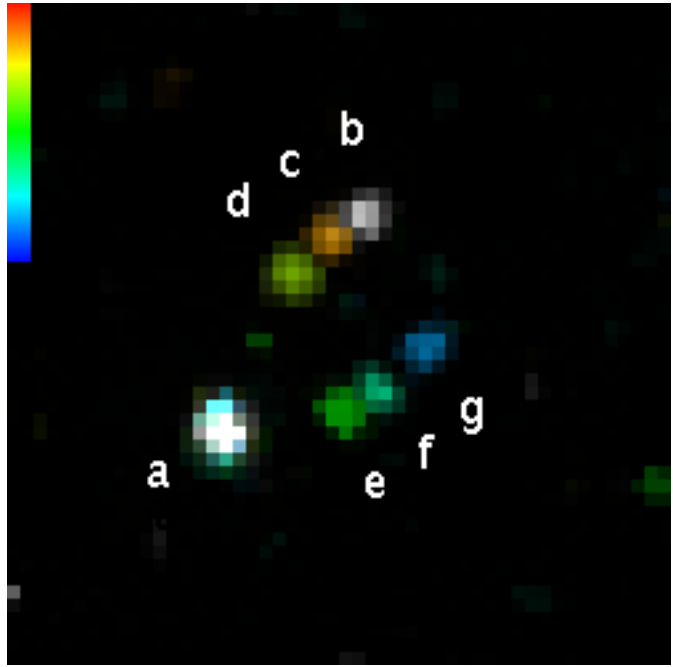

Figure 2. Images of binary TNO 1998 WW31 taken by the HST presented as a CIP. The images have been aligned so the brightest object (a) appears stationary. The second object was imaged in July (b), August (c), September (d) and December (e) of 2001 and again in January (f) and February (g) of 2002 to delineate its orbit.